\documentclass[
 twocolumn,
 amsmath,amssymb, 
 aps,
 prd,
 letterpaper,
 superscriptaddress,
 showpacs, showkeys,
floatfix,
 nofootinbib,
]{revtex4-2}%

\usepackage[colorlinks,linkcolor=blue,anchorcolor=blue, citecolor=blue,urlcolor=blue,]{hyperref}
\usepackage{graphicx}
\usepackage{color}
\usepackage{xcolor}
\usepackage{float}
\usepackage{makecell}
\usepackage{newtxtext,newtxmath}
\usepackage{fancyhdr}
\usepackage{appendix}
\usepackage{amssymb}
\usepackage{multirow}
\usepackage{booktabs}
\usepackage{subfigure}
\usepackage{natbib}
\usepackage[marginal]{footmisc}
\usepackage{tikz}
\usepackage{tabularx}
\usepackage{algorithm}
\usepackage{algorithmic}
\usepackage{xcolor}
\usepackage{aas_macros}
\usepackage{enumitem}
\usepackage{ulem}
\usepackage{footnote}
\usepackage{orcidlink}

\definecolor{revision}{RGB}{10,80,150}

\definecolor{revision}{RGB}{10,80,150}

\makeatletter
\usetikzlibrary{shapes.geometric, arrows.meta, positioning}

\begin{document}

\preprint{APS/123-QED}

\title{Lambda as a Probe of Lensing Consistency}

\author{Li Cui\,\orcidlink{0009-0007-6825-4890}}
\email{nervous-cl@sjtu.edu.cn}
\affiliation{State Key Laboratory of Dark Matter Physics, School of Physics and Astronomy, Shanghai Jiao Tong University, Shanghai 200240, China}
\affiliation{Key Laboratory for Particle Astrophysics and Cosmology (MOE) / Shanghai Key Laboratory for Particle Physics and Cosmology, China}

\author{Yuan Shi\,\orcidlink{0000-0001-8233-3703}}
\email{shiyuan0929@gmail.com}
\affiliation{State Key Laboratory of Dark Matter Physics, School of Physics and Astronomy, Shanghai Jiao Tong University, Shanghai 200240, China}
\affiliation{Key Laboratory for Particle Astrophysics and Cosmology (MOE) / Shanghai Key Laboratory for Particle Physics and Cosmology, China}
\affiliation{Shanghai Astronomical Observatory (SHAO), Nandan Road 80, Shanghai, China}

\author{Carlo Giocoli\,\orcidlink{0000-0002-9590-7961}} 
\affiliation{INAF-Osservatorio di Astrofisica e Scienza dello Spazio di Bologna, Via Piero Gobetti 93/3, I-40129 Bologna, Italy}
\affiliation{INFN – Sezione di Bologna, viale Berti Pichat 6/2, I-40127 Bologna, Italy}

\date{\today}

\begin{abstract}
We introduce a framework to identify the radial transition in mass reconstruction reliability between strong and weak gravitational lensing in galaxy clusters. In weak lensing reconstruction, the convergence recovered from the reduced shear is subject to the mass-sheet degeneracy. We demonstrate that the degeneracy itself can serve as an indicator of the reconstruction reliability, and introduce a spatially resolved parameter $\lambda(r)$ to characterize this as a function of radius. We validate this approach on simulated clusters with realistic observational noise, and show that the radial behavior of $\lambda(r)$ captures the relative reliability of the two probes. Furthermore, when the global mass-sheet parameter $\lambda$ is constrained directly using strong lensing information, the tightest constraints arise where the two probes achieve comparable precision. This provides a quantitative basis for joint strong and weak lensing mass reconstruction.

\end{abstract}

\maketitle

\section{Introduction} \label{sec:intro}
Gravitational lensing is a powerful tool for constraining matter distribution in galaxy clusters \citep{Jullo2010, Umetsu2016, clash2012, KIDS2018}. Unlike conventional methods such as optical, X-ray, or Sunyaev–Zel’dovich observations \citep{Carlberg1996, Xray2009, Planck2014, Mroczk2019}, which trace baryonic components and rely on equilibrium assumptions, lensing directly probes the spatial variations of the gravitational potential. As the most massive and spatially extended structures in the universe, galaxy clusters naturally exhibit both strong and weak gravitational lensing effects. Strong lensing (SL) arises in the dense cluster core, where multiple images or giant arcs can form, whereas weak lensing (WL) is more effective in the outskirts, producing only subtle statistical distortions. A key challenge is to synthesize the two approaches into a coherent framework while quantifying their relative strengths across different cluster regions. 

Weak lensing mass reconstructions of galaxy clusters are subject to the mass-sheet degeneracy \citep{Falco1985}. 
Under this degeneracy, the transformation $\kappa \rightarrow \kappa' = \lambda \kappa + (1 - \lambda), \gamma \rightarrow \gamma' = \lambda \gamma$ for arbitrary $\lambda \neq 0$ leaves the reduced shear field invariant. 
The mass-sheet degeneracy can be broken using strong lensing information from systems with multiple background sources at known redshifts \citep{Bradac2004,Bradac2005}. 
In practice, the factor $\lambda$ can be constrained by comparing the convergence fields from strong and weak lensing in their overlapping region. 

The mass-sheet transformation is inherently global and introduces a single degree of freedom. However, lensing reconstructions yield two-dimensional convergence maps, and the comparison between the strong and weak lensing maps is performed on a pixel-by-pixel basis over aligned grids. This provides an effective local estimate of $\lambda$ in each pixel, forming a spatially resolved $\lambda$ map. 
In radial bins where both methods yield reliable constraints, the inferred $\lambda$ most closely recovers the true value.
Conversely, deviations arise in areas where one method contributes substantially weaker constraints. 

In this work, we use simulations to study how the $\lambda(r)$ correlates with the accuracy of lensing mass reconstructions. We generate simulated clusters with realistic observational noise, including simulated strong lensing image configurations and weak lensing source galaxy densities. 
With the true convergence map known from the simulations, we test whether the radial behavior of $\lambda(r)$ correlates with the radial variation in the relative accuracy of strong and weak lensing reconstructions. When the mass-sheet transformation parameter $\lambda$ is independently determined by external probes, this allows us to determine which method provides the more accurate reconstruction at different radii. Conversely, when the $\lambda$ is inferred from the comparison of strong and weak lensing convergence maps, its radial behavior helps identify the radius at which the relative accuracy of strong and weak lensing reconstructions reverses.

The outline of the paper is as follows. In section 2, we begin with the theoretical basis of gravitational lensing. 
Section 3 presents the simulation setup for generating mock data. Section 4 details the mass mapping procedures, presents the inferred $\lambda(r)$ profiles, and compares the reconstructed convergence profiles with the underlying truth. Section 5 summarizes our findings and discusses future prospects.

Throughout this paper, we adopt a standard flat $\Lambda$CDM cosmology with $\Omega_{\mathrm{m}} = 0.3$, $\Omega_{\Lambda} = 0.7$, and $h = 0.7$. All magnitudes are expressed in the AB system \citep{Oke74}.

\section{Methodology}
In this section, we begin with the basic theory of gravitational lensing, followed by the framework of our method, which is applied to the simulations and subsequent analyses in the following sections.

The gravitational lensing transformation maps the coordinates on the source plane $\boldsymbol{\beta} = (\beta_1, \beta_2) $ to the coordinates on the lens plane $\boldsymbol{\theta} = (\theta_1, \theta_2)$ as 
\begin{equation}
    \boldsymbol{\beta} = \boldsymbol{\theta} - \nabla \psi(\boldsymbol{\theta})
\end{equation}
where $\boldsymbol{\theta}$ and $\boldsymbol{\beta}$ denote the image and source positions, and $\psi(\boldsymbol{\theta})$ is the projected lens potential computed at the image position (details see Ref.~\citep{Schneider1992}).

The convergence $\kappa$ is defined as a dimensionless surface mass density
\begin{equation}
    \kappa(\boldsymbol{\theta}) \equiv \frac{\Sigma(\boldsymbol{\theta})}{\Sigma_{\rm{cr}}} 
\end{equation}
Here $\Sigma(\boldsymbol{\theta}$) denotes the surface mass density at $\boldsymbol{\theta}$, with
\begin{align}
    \Sigma_{\rm{cr}} = \frac{c^2}{4\pi G}\frac{D_\mathrm{S}}{D_\mathrm{L} D_\mathrm{LS}}
\end{align}
is the critical surface mass density, where $D_{\mathrm{L}}$, $D_{\mathrm{S}}$ and $D_{\mathrm{LS}}$  are the angular diameter distances from the observer to lens, observer to source and lens to source, respectively. 

The distortion of the images can be described by the Jacobian matrix \citep{Massimo2021}
\begin{equation}
    A_{ij} \equiv \frac{\partial \beta_i}{\partial \theta_j} = \delta_{ij} - \frac{\partial^2 \psi(\boldsymbol\theta)}{\partial \theta_i \partial \theta_j} 
\end{equation}
The second derivatives of the lensing potential $\psi(\boldsymbol{\theta})$ 
can be decomposed into the convergence and shear components as
\begin{align}
\kappa = \frac{1}{2}(\partial_1^2 + \partial_2^2) \psi, \quad \gamma_1 = \frac{1}{2}(\partial_1^2 - \partial_2^2) \psi, \quad \gamma_2 =\partial_1\partial_2 \psi
\end{align}
Thus, the magnification factor is given by the inverse of the determinant of the Jacobian matrix
\begin{align}
    \mu = \frac{1}{\mathrm{det} \boldsymbol{A}} = \frac{1}{(1-\kappa)^2 - |\gamma|^2}  \quad \text{where}\; \gamma = \gamma_1 + i \gamma_2\
\end{align}
In the weak lensing regime, where the image distortion is small ($|\kappa|, |\gamma| \ll 1$), the lensing effect can be statistically inferred from the observed ellipticities of background galaxies. The observable quantity is the reduced shear
\begin{align}
    g(\boldsymbol{\theta}) = \frac{\gamma(\boldsymbol{\theta})}{1-\kappa(\boldsymbol \theta)}
\end{align}
which remains invariant under mass sheet transformation $\kappa \rightarrow \lambda \kappa + (1 - \lambda)$ and $\gamma \rightarrow \lambda \gamma$, indicating that different mass distributions yield identical observables in the weak lensing reconstruction.

The mass-sheet transformation can be expressed in terms of the lensing potential as
\begin{align}
    \psi'(\boldsymbol{\theta}) = \lambda \psi(\boldsymbol{\theta}) + \frac{1}{2}(1-\lambda) |\boldsymbol{\theta}|^2 
\label{potential}
\end{align}
where $\lambda$ is a global scaling factor. In strong lensing systems with multiple sources at different redshifts, all groups of images jointly constrain the same underlying lens potential and hence a common value of $\lambda$.
 
For a given image position $\boldsymbol{\theta}$, we denote the true convergence corresponding to source redshifts $z_\mathrm{s_1}$ and $z_\mathrm{s_2}$ as $\kappa_1^{\text{true}}$ and $\kappa_2^{\text{true}}$, which satisfy
\begin{align}
    \mathcal{R} \equiv \frac{\kappa_1^{\text{true}}(\boldsymbol{\theta})}{\kappa_2^{\text{true}} (\boldsymbol{\theta})} 
    =
    \frac{\Sigma_{\mathrm{cr}}(z_l, z_{\mathrm{s2}})}
         {\Sigma_{\mathrm{cr}}(z_l, z_{\mathrm{s1}})}
\label{ratio}
\end{align}
Since this ratio $\mathcal{R}$ is determined solely by the lensing geometry, it takes a unique value once the lens and source redshifts are specified. In contrast, under the mass-sheet transformation, the ratio becomes
\begin{align}
    \mathcal{R'} =
    \frac{\kappa_1'(\boldsymbol{\theta})}{\kappa_2' (\boldsymbol{\theta})}
    = \frac{\lambda \kappa_1^{\text{true}} + (1-\lambda)}{\lambda \kappa_2^{\text{true}} + (1-\lambda)} 
\label{exp_ratio}
\end{align}
which in general differs from $\mathcal{R}$.  With $\kappa_1'$ and $\kappa_2'$ obtained by taking derivatives of $\psi'$ in Eq.~(\ref{potential}), and $\mathcal{R}$ determined by the known redshifts, the scaling factor $\lambda$ is
\begin{align}
    \lambda = 1 + \frac{\kappa_1'(\boldsymbol{\theta}) - \mathcal{R} \kappa_2'(\boldsymbol{\theta})}{\mathcal{R}-1}
\end{align} 
Through Eqs.~(\ref{ratio}) and (\ref{exp_ratio}), each source pair $(z_{s_1}, z_{s_2})$ provides an independent constraint on $\lambda$. One might therefore expect $\lambda$ to be fully determined. However, the constraining power of each pair is intrinsically entangled with the uncertainties  in reconstructing the lensing potential, since the values of $\kappa_1'$ and $\kappa_2'$ are not directly observed but inferred from reconstructed  lensing potential. Consequently, regions where the strong lensing reconstruction is less reliable provide substantially  weaker constraints on $\lambda$.

\section{Simulation}
We generate the mass distribution of the galaxy cluster using $\texttt{MOKA}$ \footnote{http://cgiocoli.wordpress.com/research-interests/moka} \citep{C12a}, a semi-analytic code that constructs cluster mass distributions from three components: smooth triaxial dark matter halos, cluster member subhalos, and brightest cluster galaxies \citep{MM17}. Member galaxies are assigned to subhalos through the halo occupation distribution (HOD), with their stellar masses and B-band luminosities set according to the host subhalo mass following \citet{Wang06}. For a detailed description of the \texttt{MOKA} code, we refer the reader to \citet{C12a,C12b}. The simulated cluster consists of  two closely separated mass peaks at redshift $z=0.3$, with virial masses $M_1 = 6.75 \times 10^{14}\,h^{-1} \rm{M_{\odot}}$ and $M_2 = 5.25 \times 10^{14}\,h^{-1}\rm{M_{\odot}}$.
Their centers are separated by $\sim 60\,\mathrm{kpc}\,h^{-1}$, with a position angle difference of $\sim25 \; \text{deg}$ between the two clumps. The simulation produces a surface mass density field on an $8 \times 8 \;\rm {arcmin^2}$ grid of $4096 \times 4096$ pixels (pixel scale $\approx 0.117 \; \rm{arcsec}$ ), which is converted to a convergence map at a reference source redshift $z_s = 2.0$. 

 We adopt a cluster with a closely separated, double-peaked mass distribution in our simulations for two reasons. First, the mass complexity of such systems provides a more stringent test of the reconstruction than a relaxed single halo. Second, the strong lensing cross-section is significantly enhanced in merging clusters \citep{Torri04}, making  these systems preferential targets for lensing surveys. As a result,  such clusters are among the most data-rich systems in both strong and weak lensing \citep{Jauzac15, B23, Med16, Clowe12}, rendering them natural candidates for the multi-probe approach considered here. The $8\times8\,\mathrm{arcmin}^2$ field extends to approximately  $0.8\,\mathrm{Mpc} \, h^{-1}$ from the cluster center. Although this field of view is smaller than those of conventional weak lensing analyses, it fully covers the radial range over which the relative reliability of strong and weak lensing reconstructions transitions. The reconstruction framework in this weak lensing regime was presented and validated in our previous work \citep{Shi26}.

\subsection{Strong Lensing Simulation}
\label{sl_sim}
We generate mock strong lensing observations by constructing a background source population at redshifts up to $z_s=9.0$. Sources are distributed uniformly over a $ 30^{\prime\prime}\times30^{\prime\prime}$ angular field, which fully covers the largest caustic area at $z_s = 9.0$. The source redshifts and intrinsic UV absolute magnitudes are sampled from the joint probability distribution
\begin{align}
    P(M_{\rm{UV}}, z) = P(z) P(M_{\rm{UV}}|z)
\end{align}
where $P(z)$ is the normalized redshift distribution obtained by integrating the luminosity function over magnitude, and $P(M_{\rm{UV}}| z)$ is the corresponding conditional luminosity function at that redshift.
The UV luminosity function follows \citet{Cuc12} at $z<2$, while for $2 \le z \le 9$ we use the measurements from \citet{Bouwens21}. The expected number of source galaxies within the field of view is given by
\begin{align}
N_{\rm exp}= \Omega \int \frac{dV}{d\Omega\,dz} \left[
    \int \Phi(M,z)\,dM \right]dz 
\end{align}
where the luminosity function is integrated over the absolute UV magnitude range from the bright-end limit $M_{\rm UV}=-25.0$ to the faint-end limit set by an apparent magnitude cut $m_{\rm{cut}} = 32.0$. The resulting number of sources is randomly drawn from a Poisson distribution with mean $N_{\rm exp}$.

The generated sources are treated as point sources and ray-traced through the lens to determine whether they produce multiple images. For each source that produces multiple images, we compute the apparent magnitude of each image from its lensing magnification. We apply a position-dependent threshold to account for the reduced detectability of images due to the contamination from the brightest cluster galaxy (BCG) and intracluster light (ICL)
\begin{align}
m_{\rm lim}(r) = m_{\rm tele} - \Delta m\exp\left(-\frac{r}{r_s}\right)
\end{align}
where r is the projected distance from the cluster center. We choose $m_{\rm tele} = 29.5$ as the limiting apparent magnitude \citep{Bezan24}, with $\Delta m = 1.5$ and $r_s = 70 \, \rm{kpc} \, h^{-1}$ to describe the radial variation of the detection limit. Based on this criterion, sources with at least two images brighter than the limiting magnitude at their projected positions are identified as strong lensing systems. We apply a Gaussian perturbation of $\sigma_{\rm pos} = 0.15 ^{\prime \prime}$ to each image position to represent the positional uncertainties typically adopted in strong lensing analyses. The resulting mock catalog provides the image positions and source redshifts of all identified strong lensing systems. In addition, the B-band luminosities of cluster members with masses above $10^{10} h^{-1} \, \, M_{\odot}$, are used as inputs for the galaxy-scale component of the lensing reconstruction.

\subsection{Weak Lensing Simulation}
\label{sec:wl_sim}

We begin by deriving the reduced shear field from the simulated convergence map. The shear components $\gamma_1$ and $\gamma_2$ are obtained using the Kaiser–Squires (KS) relation \citep{Kaiser1993}. The reduced shear is then computed as 
\begin{align}
    g(\boldsymbol{\theta}) = g_1(\boldsymbol{\theta}) + i g_2(\boldsymbol{\theta}) = \frac{\gamma_1 (\boldsymbol{\theta}) + i \gamma_2(\boldsymbol{\theta})}{1-\kappa(\boldsymbol{\theta})}
\end{align}
on a regular $48 \times 48$ grid covering an $8^{\prime}\times 8^{\prime}$ field of view. Each pixel has an angular size of 
$0.167^{\prime}\times0.167^{\prime}$, equivalent to a physical scale of $31.15\,\mathrm{kpc} \, h^{-1}$ per pixel at cluster redshift of $z=0.3$. We apply a binary mask to pixels where the reduced shear $|g|$ exceeds $0.5$, consistent with the reduced shear range covered by current JWST cluster weak lensing catalogs \citep{Harvey24}. To account for the effects of lensing magnification and contamination from foreground cluster galaxies \citep{Broadhurst95, Schmidt09}, we approximate the effective galaxy number density as a function of projected cluster-centric distance $R$
\begin{align}
n_{\rm gal}(R)
    = n_0 \left( 1 - 0.3\,e^{-R/R_s} \right)
    \label{nd}
\end{align}

We adopt $R_s = 210 \, \rm{kpc} \, h^{-1}$ and $n_0 = 432 \, \rm {arcmin ^{-2}}$, yielding an average of $\sim 12$ galaxies per pixel in the outskirts of the cluster field. This source density is slightly higher than the source density reported in recent JWST weak lensing observations \citep{Cha25}. We also consider a lower source number density of $n_0 = 396 \, \rm {arcmin ^{-2}}$, with the results shown in Appendix \ref{App}.
The expected number of galaxies in each pixel is
\begin{align}
    \bar{N}_{ij} = n_{\rm gal} (R_{ij}) A_{\rm pix}
    \label{eq:Nij}
\end{align}
where $R_{ij}$ is the projected distance from the cluster center to the center of pixel $(i, j)$ and $A_{\rm pix}$ is the pixel area. The galaxy count in each pixel used in the mock catalog is then drawn from a Poisson distribution with mean $\bar{N}_{ij}$. 

\begin{figure}
    \centering
     \includegraphics[width=0.5\textwidth]{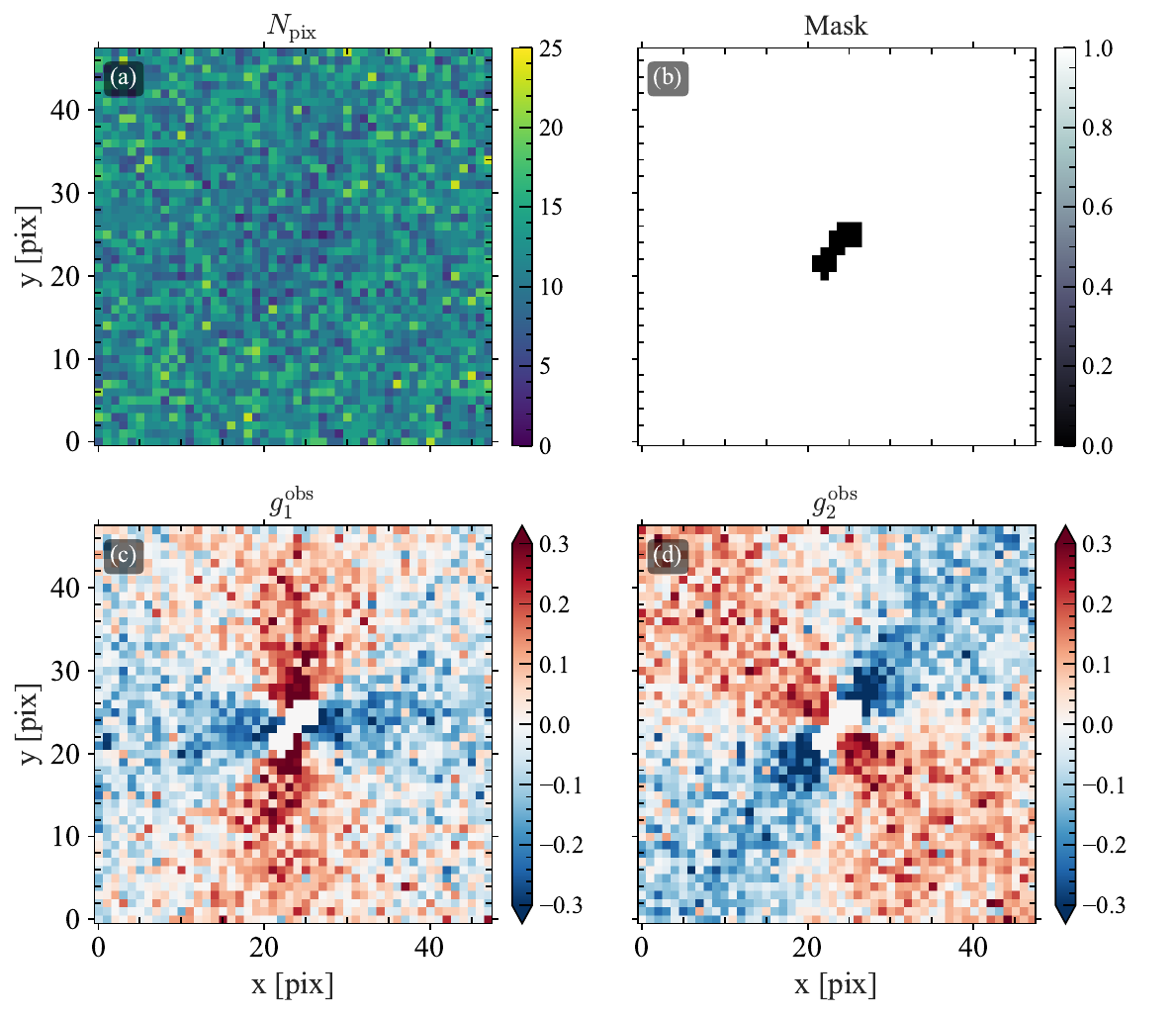}
     \caption{Mock weak lensing data on the $48\times48$ pixel grid.
    From left to right: the number of background galaxies per pixel, $N_{\rm pix}$; the binary mask (1 for included pixels and 0 for masked pixels); the two simulated shear components $\hat{g}_1$ and $\hat{g}_2$ (Eq.~(\ref{eq:ghat})).}
    \label{fig:input_data}
\end{figure}

Each source galaxy is first assigned an intrinsic complex ellipticity $\varepsilon^{\rm s}=\varepsilon^{\rm s}_1+i\varepsilon^{\rm s}_2$, with each component drawn independently from a Gaussian distribution with zero mean and standard deviation $\sigma_\varepsilon=0.24$ \citep{B01, Lea07}. The observed ellipticity in the $|g|\le1$ regime is obtained from the local reduced shear using the transformation  \citep{SeitzSchneider1997}
\begin{equation}
\varepsilon \;=\; \frac{\varepsilon^{\rm s}+g}{1+g^{*}\,\varepsilon^{\rm s}},
\label{eq:eps_transform}
\end{equation}
where $g_*$ denotes the complex conjugate of $g$. The reduced shear is estimated in each pixel by averaging the observed ellipticities of the $N_{\rm pix}$ galaxies in that pixel
\begin{equation}
\hat g = \frac{1}{N_{\rm pix}}\sum_{k=1}^{N_{\rm pix}}
\varepsilon_k ,
\label{eq:ghat}
\end{equation}
This estimator is unbiased for an isotropic intrinsic-ellipticity distribution, i.e., $\langle \hat{g} \rangle = g$. Since the reduced shear can be comparable to unity in the inner regions of galaxy clusters, the approximation $\hat{g} \approx g + \sigma_{\varepsilon}/{\sqrt{N_{\rm pix}}}$ breaks down \footnote{%
In the regime where $|g|\lesssim0.2$, the shape noise in each pixel is well approximated by a Gaussian distribution with variance $\sigma_{\rm pix}^2=\sigma_\varepsilon^2/N_{\rm pix}$.}. Therefore, $\hat{g}$ is estimated following the procedure described above. The resulting background galaxy distribution, binary mask, and mock reduced shear maps are shown in Fig.~\ref{fig:input_data}.

\section{Results}
\label{total_result}
In this section, we first introduce the strong lensing mock data generated following the procedure described in Section~\ref{sl_sim}. We then describe the strong lensing modeling procedure and assess the performance of the resulting model. Subsequently, we detail the methodology we used in weak lensing mass reconstruction. We then discuss the resulting mass maps and their reconstruction accuracy. Finally, we compare the reconstructed convergence maps from strong and weak lensing on a pixel-by-pixel basis, from which we derive the radial profile of the mass-sheet degeneracy parameter $\lambda(r)$. These results suggest that $\lambda(r)$ can serve as a useful indicator for combining strong and weak lensing constraints in full-field mass reconstruction. 

\subsection{Strong Lensing Reconstruction}
\label{subsec:sl_recon}

From the simulated strong lensing mock observations, we identify 22 multiply imaged systems comprising 68 individual images, which are used as positional constraints in the strong lensing reconstruction. We perform the reconstruction with \texttt{GLAFIC} \citep{Oguri10}, assuming a single lens plane and using only the multiple-image positions as observables. 

The total mass distribution of the cluster is modeled as the sum of two main mass clumps and a cluster member galaxy component. Each mass clump consists of a halo component modeled by an elliptical Navarro-Frenk-White (NFW) mass density profile \citep{Na96} and a BCG component modeled by an elliptical Hernquist mass density profile \citep{Hern90}, with their lensing properties calculated using the fast approximation proposed by \citet{Oguri2021}. Each cluster member galaxy is modeled with the dual pseudo-isothermal elliptical (dPIE; \texttt{gals} in \texttt{glafic} mass profile) \citep{Jaffe83, Keeton01}, with its ellipticity fixed to zero for simplicity. The total mass of each member galaxy scales as $M_{\rm{tot},i} \propto \sigma_{v,i}^2 r_{\rm{cut}, i}$. To reduce the number of free parameters, we assume that the velocity dispersion and truncation radius follow the luminosity scaling relations:
\begin{align}
    \sigma_i = \sigma^{\rm{ref}} (\frac{L_i}{L_{\rm{ref}}})^{\alpha} \; , \;
    r_{\rm{cut}, i} = r^{\rm{ref}}_{\rm{cut}} (\frac{L_i}{L_{\rm{ref}}})^{\beta}
    \label{scaling_eqn}
\end{align}
where $\alpha = 0.25$ and $\beta = 0.5$, and $L_i$ is the $B$-band luminosity of the $i$-th galaxy in the mock catalog generated by \texttt{MOKA}.

For each cluster-scale halo, we adopt an NFW profile with the mass, centroid position, ellipticity, position angle, and concentration treated as free parameters. The mass is optimized over the range of $10^{14}$-$10^{15}\,h^{-1}\,M_{\odot}$, while the concentration and ellipticity are allowed to vary over the ranges of $1-6$ and $0.1-0.7$, respectively. We model each BCG with a Hernquist profile and the mass, centroid position, ellipticity, position angle, and scale radius are included as free parameters, with the total mass optimized over the range of $8\times10^{11}-8\times10^{12}\,h^{-1}\,M_{\odot}$ and the ellipticity over the range of $0.1-0.7$. The halo and BCG centroids are initialized at the approximate observed positions of the corresponding BCGs and allowed to vary within $\pm5''$ during the optimization.

For the cluster member galaxy component, we include member galaxies within a rectangular region enclosing all multiple images, with a $10^{\prime \prime}$ padding added to each boundary. Since variations in the properties of more distant member galaxies produce minor changes in the predicted image positions, they are weakly constrained by the strong lensing data and are therefore excluded from the optimization. We apply the scaling relations in Eq.~(\ref{scaling_eqn}), taking the brightest cluster member excluding the BCGs as the reference galaxy. The normalization parameter $\sigma^{\rm ref}$ is optimized over $80$--$300\,\mathrm{km\,s^{-1}}$, while $r_{\rm cut}^{\rm ref}$ is optimized over $1''$--$50''$. 

Additionally, since the mock cluster does not include an external shear component, we do not include this in our lens model. As a consistency check, we tested models including an external shear component and found that, although it reduced the convergence map residuals in localized regions, the overall reconstruction became less accurate. In the case of real cluster lenses, this suggests that the inclusion of external shear and other higher-order perturbations should be carefully evaluated \citep{Amruth26}, as they may not necessarily improve the recovery of the underlying mass distribution.    

We optimize the lens model parameters via standard $\chi^2$ minimization implemented in \texttt{glafic}. The $\chi^2$ is evaluated in the source plane to accelerate the optimization (see \citet{Oguri10} for details). The optimized lens model is then used to predict the positions of the multiple images, from which we compute the root-mean-square (RMS) offset between the observed and predicted image positions 
\begin{align}
    \Delta \boldsymbol{\theta}_{\rm {rms}} = \sqrt{\frac{1}{N_{\rm img}} \sum_{i=1}^{N_{\rm img}}|\boldsymbol \theta_i^{\rm{obs}} -\boldsymbol{\theta}_i^{\rm pred}|^2}
\end{align}
where $N_{\rm img} = 68$ is the total number of multiple images, $\boldsymbol{\theta}_i^{\rm obs}$ and $\boldsymbol{\theta}_i^{\rm pred}$ are the observed and predicted image positions. The best-fit model yields $\Delta \boldsymbol{\theta}_{\rm rms} = 0.186^ {\prime\prime}$ and is adopted for the subsequent analysis. We further access the model by computing reduced-$\chi^2$, which is defined as
\begin{align}
    \chi^2_{\nu} = \frac{1}{2N_{\rm img} - P} \sum _{i=1} ^{N_{\rm img}} \frac{|\boldsymbol \theta_i^{\rm{obs}} -\boldsymbol{\theta}_i^{\rm pred}|^2}{\sigma_{\rm pos}^2}
\end{align}
where total observational constraints is $2N_{\rm img} = 136$, $P = 26$ is the number of free parameters, and $\sigma_{\rm pos} = 0.15^{\prime \prime}$ is the positional uncertainty adopted in the mock image generation. Since the same positional uncertainty is assumed in both coordinate directions, the two-dimensional positional residual is adopted in the above expression. For our best-fit model, this yields $\chi^2_{\nu} \approx 0.95$.

\subsection{Weak Lensing Reconstruction}
\label{subsec:wl_recon}
\begin{figure}
    \centering
    \includegraphics[width=0.5\textwidth]{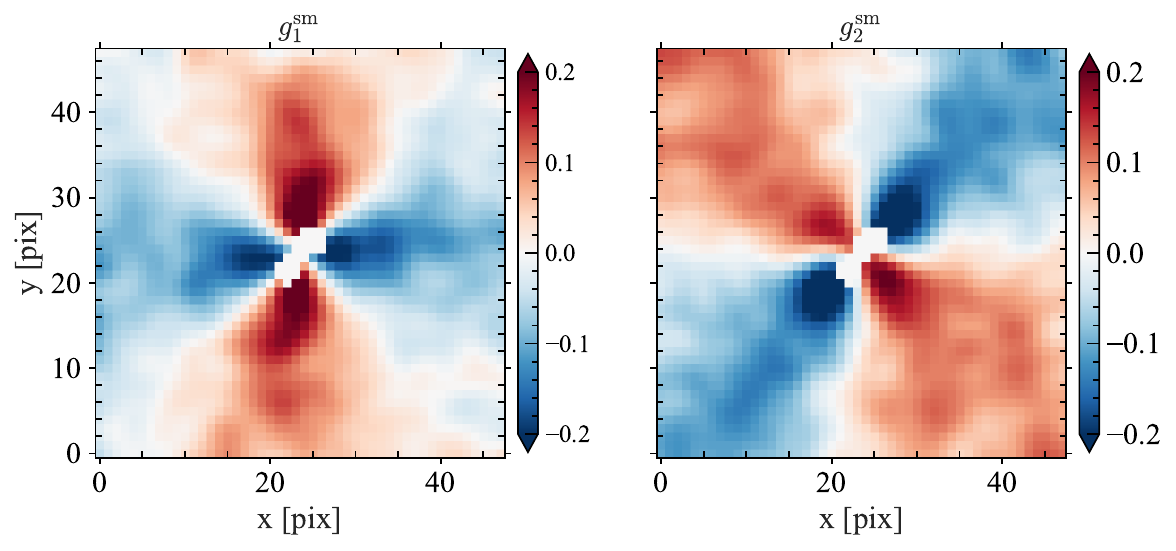}
    \caption{Smoothed reduced shear components $g_1^{\rm sm}$ (left) and
    $g_2^{\rm sm}$ (right) used as input to the weak lensing
    reconstruction.}
    \label{fig:input_data_sm}
\end{figure}

We begin by smoothing the reduced shear field with a Gaussian kernel before performing the mass reconstruction, since individual galaxy ellipticity measurements are dominated by shape noise. For a Gaussian kernel with width $\theta_{\rm sm}$, the smoothed reduced shear $g^{\rm sm}_{1,2}$ is given by
\begin{equation}
    g^{\rm sm}_{1,2} =
    \begin{cases}
    \dfrac{(g_{1,2}\, w) \otimes S_{\theta_{\rm sm}}}
          {w \otimes S_{\theta_{\rm sm}}}
      & \text{if } w \otimes S_{\theta_{\rm sm}} > 0 \\[8pt]
    0 & \text{if } w \otimes S_{\theta_{\rm sm}} = 0 
    \end{cases}
    \label{eq:smoothing}
\end{equation}
where $S_{\theta_{\rm sm}}$ is the Gaussian kernel and $w$ is the weight function. In general $w$ incorporates the observational weights from individual galaxy shape measurements. Here we adopt a flat observational weighting scheme for simplicity; consequently, $w$ is entirely determined by the mask. This reduces $w$ to a binary map where $w=1$ for unmasked pixels containing valid data and $w=0$ for masked or empty regions.
We adopt $\theta_{\rm sm} =1.5$ pixels at the map resolution of $0.167'$ per pixel.  The mock reduced shear field is generated assuming a single source plane. This simplification does not account for source redshift uncertainties, and the resulting reconstruction is based on an idealized setup. The smoothed components $g_1^{\rm sm}$ and $g_2^{\rm sm}$ are shown in Fig.~\ref{fig:input_data_sm}. We then zero pad this smoothed field from $48 \times 48$ to $96 \times 96$, as the subsequent Fourier transform used in the mass reconstruction requires periodic boundary conditions.

\begin{figure}
    \centering
    \includegraphics[width=0.48\textwidth]{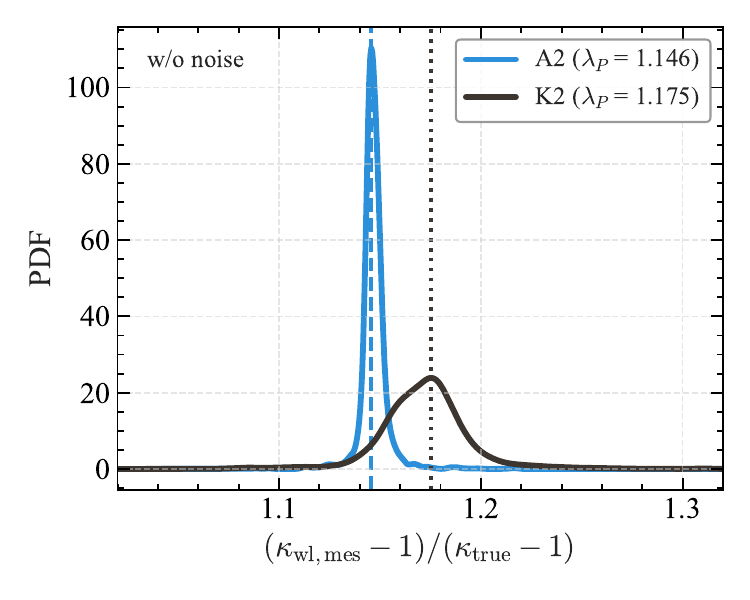}
    \caption{Probability density distribution of the effective mass-sheet parameter $\lambda(r)$ for the A2 (blue) and K2 (black) reconstruction methods. The input reduced shear field is noiseless, while the region $|g| > 0.5$ is assumed to be masked. The distribution is evaluated over unmasked pixels in the radial range $100 \le r < 700~\mathrm{kpc}\,h^{-1}$ at the fifth iteration. The dashed and dotted vertical lines indicate the peak values of the A2 and K2 distributions, with the dashed line marking A2 and the dotted line marking K2.}
    \label{fig:lambda_pdf_noise_free}
\end{figure}

We reconstruct the convergence field using the AKRA-based iterative framework developed in \citet{Shi26}, which employs a prior-free maximum-likelihood estimator to iteratively reconstruct the convergence from the reduced shear. We refer the reader to that work for a full description of the five reconstruction methods (K1, K2, A1, A2, and A3), as well as the dependence of the reconstruction error on the iteration number for the different reconstruction methods. Here, K and A denote the KS-based and AKRA-based reconstruction methods,  while the number denote the corresponding modifications introduced: 1 for the standard reconstruction, 2 for using an SIS profile as the initial guess, and 3 for adopting the smooth-mask scheme.

The reconstructed weak lensing convergence map is related to the true convergence map through a mass-sheet transformation. To characterize this transformation across the cluster field, we define an effective mass-sheet parameter $\lambda(r)$  at each pixel

\begin{align}
    \lambda(r) = \frac{\kappa_{\rm wl,\, mes}(r) -1}{\kappa_{\rm true}(r) - 1}
    \label{eq:lambda_cp_true}
\end{align}
where the $\kappa_{\rm wl,\, mes}(r)$ is the reconstructed convergence and $\kappa_{\rm true}(r)$ is the true convergence from the simulation.  The inferred $\lambda(r)$ is potentially affected by the mass-sheet transformation as well as by the reconstruction biases, with these two effects being coupled in the weak lensing reconstruction. We first examine the resulting $\lambda(r)$ distributions for the A2 and K2 methods in the noiseless case, as shown in Fig.~\ref{fig:lambda_pdf_noise_free}. For A2, the noiseless $\lambda(r)$ distribution is sharply peaked and nearly a delta function, centered on $\lambda_P = 1.146$.  This indicates that the relation described by Eq.~\ref{eq:lambda_cp_true} between the reconstructed and true convergence fields can be characterized by a single, spatially uniform scaling factor. As our analysis is based on a finite field of view, the inferred value of $\lambda_P$ can depend on the adopted field size. In contrast, the K2 distribution is broader and its peak is shifted to $\lambda_P = 1.175$. Even in the absence of shape noise, the KS-based reconstruction introduces spatially varying systematic errors that cannot be described by a single global value $\lambda$. 

\begin{figure}
    \centering
    \includegraphics[width=0.48\textwidth]{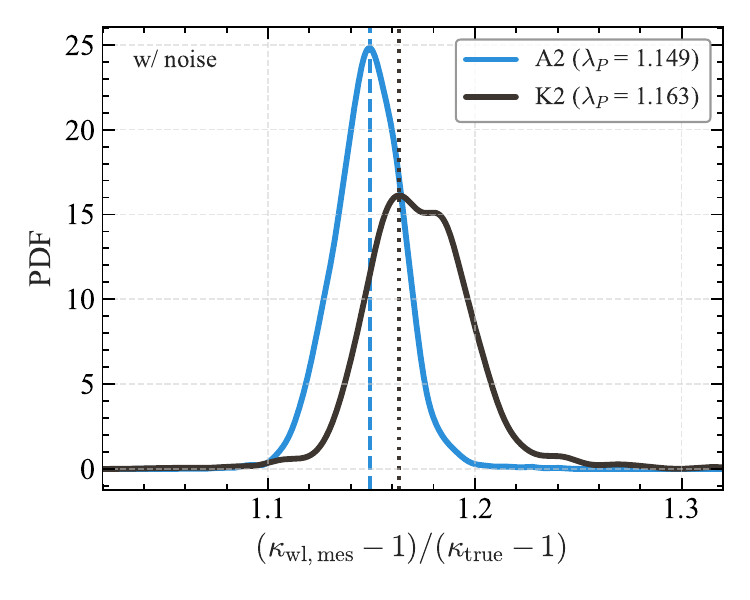}
    \caption{Same as Fig.~\ref{fig:lambda_pdf_noise_free}, but for an input reduced shear field with added shape noise. The blue curves show the results obtained with the A2 method, while the black curves correspond to the K2 method. }
    \label{fig:lambda_pdf_noise}
\end{figure}

We extend our analysis to the case with shape noise by applying the A2 and K2 methods to the mock reduced shear maps described in Section~\ref{sec:wl_sim}. As shown in Fig.~\ref{fig:lambda_pdf_noise}, the A2 distribution broadens into an approximately Gaussian profile, while its peak remains nearly unchanged compared with the noiseless case. The recovered $\lambda(r)$ profile acquires random scatter due to the presence of the shape noise. The K2 distribution spreads further, with its peak shifting to $\lambda_P = 1.163$ and exhibits a visibly asymmetric profile. In this case, the K2 reconstruction exhibits spatially varying systematic errors in addition to the noise-induced scatter. In the subsequent combined analysis, these reconstruction errors would become entangled with the assessment of the relative reliability between the strong and weak lensing reconstructions. We therefore adopt the A2 method hereafter.

\begin{figure}
    \centering
    \includegraphics[width=0.48\textwidth]{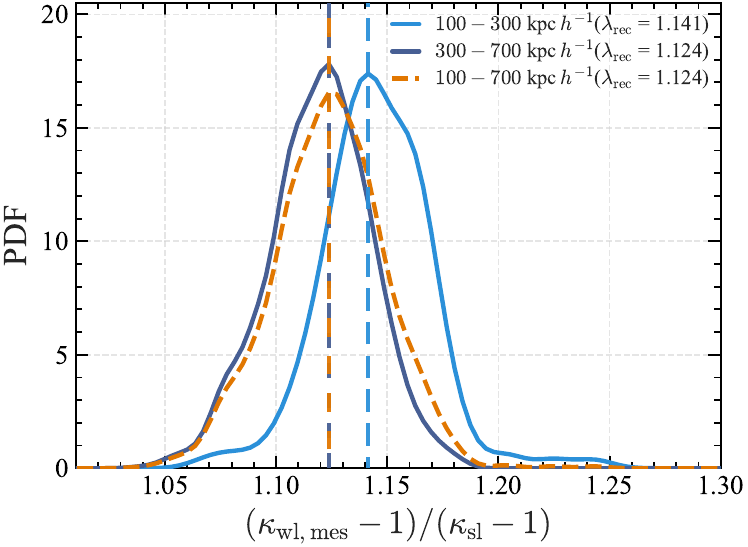}
    \caption{Probability density distributions of $\lambda_{\rm rec}$ in different radial ranges, computed from Eq.~(\ref{lambda_sl_rt}). The blue solid, dark blue solid and orange dashed curves corresponds to $100 \le r < 300$, $300 \le r < 700$, and $100 \le r < 700 \rm{kpc} \, h^{-1}$, respectively. The vertical lines mark the corresponding distribution peaks at $\lambda_P = 1.141$, $1.124$, $1.124$. When the combined radial range is used, the peak shifts away from the noiseless value of $\lambda_P = 1.146$,  suggesting that using full radial range can lead to a deviation in the inferred $\lambda$.}
    \label{fig:lambda_sl}
\end{figure}

\begin{figure*}
    \centering
    \includegraphics[width=0.68\textwidth]{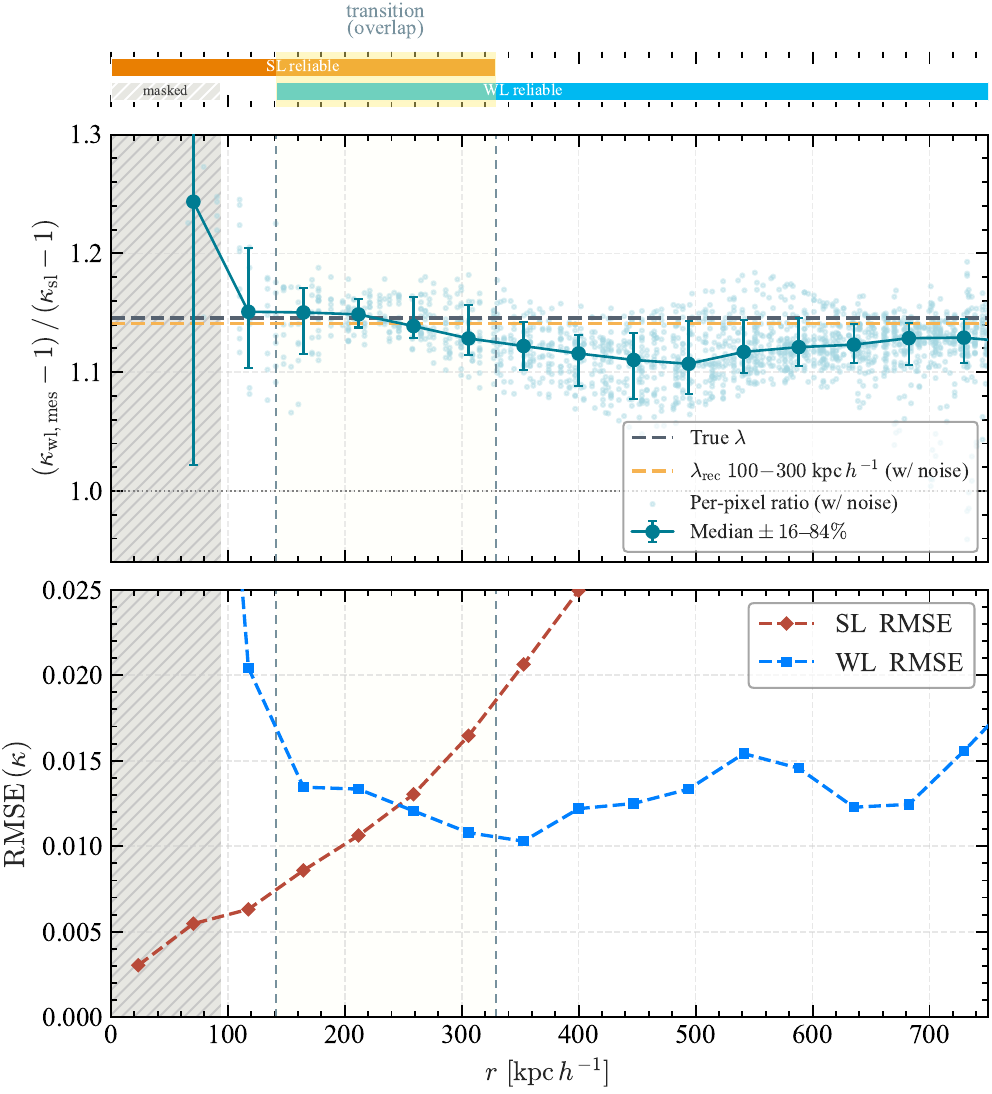}
    \caption{The upper panel shows how $\lambda_{\rm rec}(r)$ varies with radius. The horizontal black dashed line marks the true $\lambda$ value inferred from the noiseless case. The orange dashed line is the $\lambda$ value inferred from the strong and weak lensing comparison over $100$-$300\rm{kpc}\,h^{-1}$. The data points show the median $\lambda_{\rm rec}$ values in each radial bin, with error bars indicating the $16$th to $84$th percentile range. The lower panel shows the RMSE of the strong and weak lensing reconstructions. For visual guidance, the yellow shaded region marks the  region where the relative reconstruction performance of the two probes becomes comparable, defined as the radial range satisfying $|\rm{RMSE_{SL}}(r) - \rm{RMSE_{WL}}(r)| / \frac{1}{2} |\rm{RMSE_{SL}}(r) + \rm{RMSE_{WL}}(r)| < 0.5$.  The choice of 0.5 here is not physically motivated and serves only as a reference. The red and blue curves represent the RMSE of the SL and WL reconstructions, respectively.}
    \label{fig:rmse_ratio}
\end{figure*}

\subsection{Combined Analysis}
We now compare strong lensing and weak lensing reconstructed convergence maps. We define the effective mass-sheet parameter from the comparison of the two reconstructions as:
\begin{align}
    \lambda_{\rm rec}(r) = \frac{\kappa_{\rm wl, mes} (r) - 1}{\kappa_{\rm sl} (r) - 1}
    \label{lambda_sl_rt}
\end{align}
which can be decomposed into:
\begin{align}
    \lambda_{\rm rec}(r) = \frac{\kappa_{\rm wl,\, mes}(r) -1}{\kappa_{\rm true}(r) - 1} \cdot  \frac{\kappa_{\rm true}(r) - 1}{\kappa_{\rm sl}(r) - 1}
    \label{lambda_sl_eq}
\end{align}
where the first factor is the effective mass-sheet parameter $\lambda(r)$ defined in Eq.~(\ref{eq:lambda_cp_true}), while the second factor characterizes the reliability of the strong lensing reconstruction.  The comparison between strong and weak lensing reconstructions is performed over the unmasked pixels, with the $|g|>0.5$ mask used in the mock data generation. This keeps $\kappa$ away from unity, ensuring that $\lambda_{\rm rec}$ remains nonsingular. As shown in the previous section, the distribution of the first factor is centered on a global value for the A2 reconstruction, while its scatter is determined by the shape noise level, which arises from the radial variation in the background galaxy number density. The second factor varies with radius as the constraining power of the strong lensing observables decreases toward larger radii. In regions where multiple-image systems at different source redshifts are observed, the convergence is well constrained and this factor remains close to unity. At larger radii where such systems are sparse or absent, the reconstruction becomes less reliable and this factor deviates progressively from unity.

To  investigate the radial variation in the relative reliability of strong and weak lensing reconstructions, we quantify the radial dependence of $\lambda_{\rm rec}(r)$ across the cluster. As the relative constraining power of the two methods changes with radius, a single probability density distribution of $\lambda_{\rm mec}$ cannot fully capture the scale-dependent behavior of the comparison. We therefore characterize the radial transition in reconstruction reliability through the radial profile of $\lambda_{\rm mes}(r)$, shown in the upper panel of Fig.~\ref{fig:rmse_ratio}. The corresponding probability density distributions in individual radial bins are presented in Fig.~\ref{fig:lambda_sl}.

The radial profile of $\lambda_{\rm rec}(r)$ can be obtained without knowledge of the true convergence and is therefore directly applicable to real observations. As illustrated in Fig.~\ref{fig:rmse_ratio}, its behavior can be divided into several regimes. In the innermost region, the background galaxy density available for the weak lensing reconstruction is low, leading to increased shape-noise uncertainty and consequently a large scatter in $\lambda_{\rm rec}$. The small number of pixels in these bins further reduces the statistical precision of the $\lambda_{\rm rec}$ distribution. At intermediate radii, the increasing galaxy density reduces shape-noise uncertainty while the strong lensing reconstruction remains reliable,  resulting in relatively small scatter in $\lambda_{\rm rec}$. The  region where the relative constraining power of the strong and weak lensing reconstructions  becomes comparable is expected to lie within this radial interval, bounded by the last bin in which the scatter is still decreasing and the first bin in which it begins to rise. This interpretation is corroborated by both the reconstruction error statistics and the comparison between the reconstructed and true convergence profiles presented below.

 Beyond the regime where the two methods yield comparable performance, the strong lensing reconstruction becomes increasingly unreliable, leading to a systematic deviation of the second factor in Eq.~(\ref{lambda_sl_eq}) from unity and an increase in its scatter.  Since the strong lensing constraints are mainly provided by multiple images in the inner regions, the parametric lens model can extrapolate these inner region constraints to larger radii based on the assumed mass profiles and scaling relations. Deviations of the reconstructed $\kappa_{\rm sl}(r)$ from the true convergence are therefore expected at larger radii. In the outermost radial bins, the scatter of $(\kappa_{\rm wl, mes} - 1 ) / (\kappa_{\rm true} - 1)$  remains comparable to that in the preceding bins, as the impact of shape noise is nearly unchanged. Meanwhile, $(\kappa_{\rm true}-1)/(\kappa_{\rm sl}-1)$ exhibits a decreasing deviation from unity as the convergence decreases.

\begin{figure}
    \centering
    \includegraphics[width=0.48\textwidth]{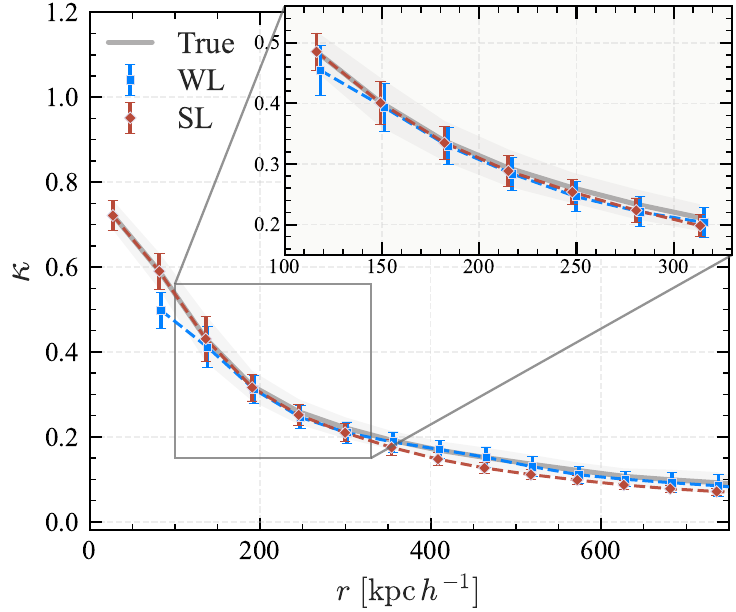}
    \caption{Radial profiles of the convergence $\kappa$ at $z_{\rm s}=2.0$ for the simulated cluster. The grey line shows the input truth, blue squares show the weak lensing reconstruction after correcting the mass-sheet degeneracy using the SL-inferred $\lambda_P = 1.141$, and red diamonds show the strong lensing reconstruction. The truth and SL profiles are smoothed to match the resolution of the WL reconstruction. The error bars indicate the scatter of $\kappa$ within each radial bin, and the shaded region around the truth profile shows its $\pm1$ standard deviation. The inset zooms into the radial range $100 \le r \le 330~\mathrm{kpc}\, h^{-1}$, where the relative accuracy of the SL and WL reconstructions is comparable.}
    \label{fig:kappa_combined}
\end{figure}

We further illustrate how the transition in the relative reliability of the two lensing probes, identified from the radial behavior of $\lambda_{\rm rec}$, relates to the underlying reconstruction accuracy. We compare it with the radial root-mean-square error (RMSE) of the reconstructed convergence, which can be quantified in simulations where the true convergence is known. The pixels are grouped into concentric annuli $A(r)$ according to their projected distance from the cluster center, and the radial RMSE is computed as:
\begin{align}
    \mathrm{RMSE}(r) = \sqrt{\big\langle\, [\kappa_{\rm rec}(\boldsymbol{\theta}) - \kappa_{\rm true}(\boldsymbol{\theta})]^2 \,\big\rangle_{\boldsymbol{\theta}\in A(r)}} , \label{eq:rmse}
\end{align}
As shown in the bottom panel of Fig.~\ref{fig:rmse_ratio}, the RMSEs of the strong and weak lensing reconstructions become comparable within the region identified from the radial behavior of $\lambda_{\rm rec}$, indicating that this region  is associated with a transition in their relative reconstruction reliability. Moreover, the estimated peak value $\lambda_P$ within this radial range differs from the true value by less than $1\%$ (see Fig.~\ref{fig:lambda_sl}), confirming that mass-sheet parameter can be recovered with high accuracy.

Finally, we apply the $\lambda$ value estimated from the strong lensing reconstruction to correct the weak lensing convergence map, and compare the resulting profile with the strong lensing reconstruction and the true convergence. As shown in Fig.~\ref{fig:kappa_combined}, the strong lensing reconstruction follows the true profile more closely in the inner region, whereas the corrected weak lensing reconstruction provides a more accurate reconstruction at larger radii. The transition between these two regimes is consistent with the radial behavior of $\lambda_{\rm rec}$ and  is supported by the RMSE analysis. Together, the two reconstructions provide complementary constraints across the full radial range, enabling a more complete reconstruction of the cluster convergence field.

\begin{figure*}
    \centering
    \includegraphics[width=0.68\textwidth]{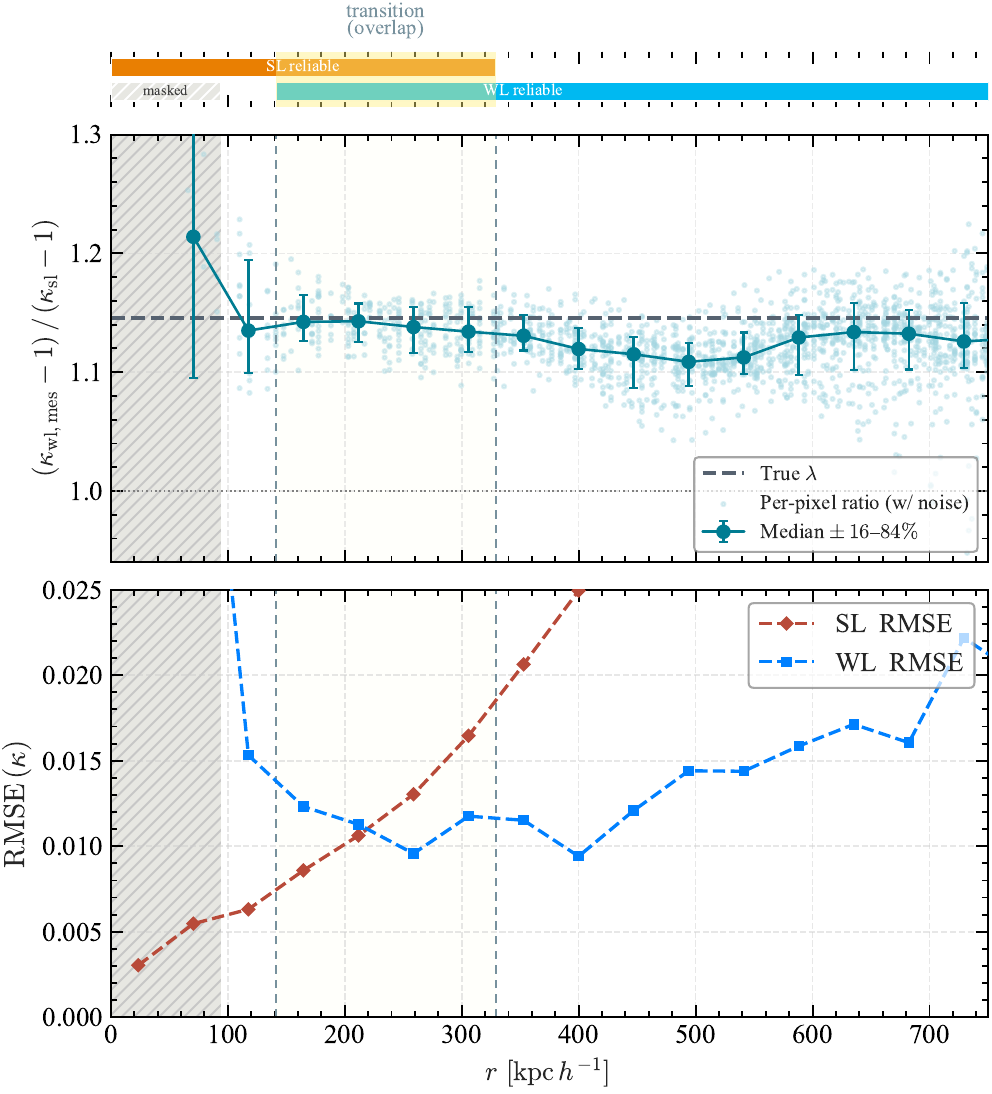}
    \caption{The upper panel shows the radial variation of $\lambda_{\rm {rec}}$, the lower panel shows the radial variation of the RMSE for the strong and weak lensing reconstructions, with $n_0 = 396\, \rm {arcmin ^{-2}}$. The panel layout and labeling follow the same convention as in Fig.~\ref{fig:rmse_ratio}.}
    \label{fig:rmse_ratio_396}
\end{figure*} 

\section{Discussion}
In this work, we investigate the relation between the effective mass-sheet transformation parameter $\lambda_{\rm rec}$, and the relative performance of strong and weak lensing reconstructions in simulated galaxy clusters. By directly comparing our reconstructed convergence maps against the known simulation truth, we establish a practical framework for assessing the consistency and reliability of these two complementary lensing probes. These results allow us to identify which of the two reconstruction methods is more reliable at specific cluster radii. Furthermore,  our results suggest that the inference of $\lambda_{\rm rec}$ from the strong lensing reconstruction is most reliable within the radial regime where both methods yield comparable precision. 

Weak lensing mass reconstruction is traditionally performed using either parametric or non-parametric methods, with machine learning approaches recently emerging as a powerful alternative. Parametric methods describe the mass distribution in terms of physically meaningful model parameters, but the resulting reconstruction is inevitably coupled to the assumed parameterization and model assumptions. For the present study, we adopt the AKRA method, a non-parametric reconstruction approach whose intrinsic reconstruction accuracy has been quantified in the noiseless iterative case, while the impact of observational masks and shape noise has been evaluated in our previous work and the present analysis. This enables the radial behavior of $\lambda_{\rm rec}$ to be interpreted in terms of the relative reliability of the two lensing probes, rather than being affected by systematic biases introduced by the reconstruction method. 

We also verified that the identified transition region is insensitive to the specific masking criterion adopted. Applying a mask to regions with $|g|>0.7$ leaves the radial behavior of $\lambda_{\rm rec}$ unchanged, suggesting that the inferred relative reliability of the two lensing probes is not significantly affected by the choice of shear cut. With the increasing number density of background galaxies and the growing number of strong lensing constraints expected from next-generation surveys, this framework may become a useful approach for studying the relative reliability of the two lensing probes and providing insights into cluster mass reconstruction.

We will extend the present framework to real observational data in future work. Such analyses will require careful treatment of additional observational effects, including photometric redshift uncertainties of background sources and other observational systematics. Broadly, how information from multiple observational tracers can be most effectively combined to study galaxy clusters remains an open question. Since the constraining power of different observational tracers varies with cluster mass and dynamical state, their optimal combination is expected to vary across the cluster population. Developing consistent frameworks to jointly exploit these complementary constraints will advance our understanding of galaxy clusters.

\section*{Acknowledgements}
We thank the anonymous referee for the constructive comments and suggestions. YS acknowledges the support from NSFC Grant No. 12503004.
This work is also supported by the National Key R\&D Program of China (2023YFA1607800, 2023YFA1607801), and the Fundamental Research Funds for the Central Universities. 
This work made use of the Gravity Supercomputer at the Department of Astronomy, Shanghai Jiao Tong University.

\appendix
\section{Additional Results}
\label{App}

To assess our method under a background source density more consistent with observations, we set $n_0 = 396\, \rm {arcmin ^{-2}}$ in Eq.~(\ref{nd}) when generating mock reduced shear data. All other simulation settings are identical to those used for the previous mock data. We then apply the same reconstruction procedure as described in Section~\ref{total_result} and show the corresponding results in Fig.~\ref{fig:rmse_ratio_396}.

At lower source density, the resulting $\lambda_{\rm rec}(r)$ appears to show larger deviations from the true $\lambda$ and increased scatter. The results shown in Fig.~\ref{fig:rmse_ratio} and Fig.~\ref{fig:rmse_ratio_396} are based on a single noise realization for each source density configuration. The increased shape noise at lower source density is expected to produce larger realization-to-realization variations, which may lead to greater variation in the reconstructed results across different realizations. These trends should be interpreted in light of this limitation.

\bibliographystyle{apsrev}
\bibliography{apssamp}
\end{document}